\documentclass[11pt]{article}
\def\half{\frac{1}{2}}

\newfont{\bbbold}{msbm10 scaled \magstep1}

\def\cA{{\cal A}}

\def\cD{{\cal D}}

\def\cF{{\cal F}}

\def\cL{{\cal L}}

\newfont{\goth}{eufm10 scaled \magstep1}

\def\c{\gamma}
\def\d{\delta}
\def\e{\epsilon}

\def\h{\eta}
\def\i{\iota}

\def\L{\Lambda}
\def\m{\mu}
\def\n{\nu}

\def\r{\rho}
\def\s{\sigma}

\def\beq{\begin{equation}}\def\eeq{\end{equation}}
\def\beqa{\begin{eqnarray}}\def\eeqa{\end{eqnarray}}
\def\barr{\begin{array}}\def\earr{\end{array}}

\def\o{\omega}
\def\del{\partial}

\def\unm{\underline m}\def\unM{\underline M}
\def\unn{\underline n}

\def\wt{\widetilde}
\def\sdet{\rm sdet}
\def\xz{\times}

\def\tE{\tilde{E}}

\def\he{\widehat{e}}
\def\hf{\widehat{f}}
\def\hv{\widehat{v}}
\def\hC{\widehat{C}}

\def\hM{\widehat{M}}

\def\ghL{\widehat{\L}}
\def\gho{\widehat{\omega}}
\def\tb{\widetilde{b}}\def\tC{\widetilde{C}}
\def\tB{\widetilde{B}}\def\tE{\widetilde{E}}

\let\la=\label

\def\nn{\nonumber}
\def\bd{\begin{document}}
\def\ed{\end{document}}
\def\ba{\begin{array}}
\def\ea{\end{array}}
\def\bea{\begin{eqnarray}}
\def\eea{\end{eqnarray}}
\def\ft#1#2{{\textstyle{{\scriptstyle #1}\over {\scriptstyle #2}}}}
\def\fft#1#2{{#1 \over #2}}
\newcommand{\be}{\begin{equation}}
\newcommand{\ee}{\end{equation}}
\newcommand{\eq}[1]{(\ref{#1})}
\newcommand{\w}[1]{\\[0.#1cm]}
\def\eqs#1#2{(\ref{#1}-\ref{#2})}
\def\det{{\rm det\,}}
\def\tr{{\rm tr}}
\newcommand{\Section}[1]{\section{#1} \setcounter{equation}{0}}
\newcommand{\hoch}[1]{$\, ^{#1}$}
\newcommand{\tamphys}{\it\small Center for Theoretical Physics,
Texas A\&M University, College Station, TX 77843, USA}
\newcommand{\kings}
{\it\small Department of Mathematics, King's College, London, UK}
\newcommand{\uu}
{\it\small Department of Theoretical Physics, Uppsala, Sweden}
\newcommand{\hip}
{\it\small HIP-Helsinki Institute of Physics, P.O. Box 64 FIN-00014 University of Helsinki,
Suomi-Finland}
\newcommand{\stock}
{\it\small Department of Theoretical Physics, Stockholm University, Sweden} \makeatletter
\renewcommand\theequation{\thesection.\arabic{equation}}
\@addtoreset{equation}{section} \makeatother

\newcommand{\auth}
{\large P.S. Howe\hoch{1}, U. Lindstr\"om\hoch{2,3} and L. Wulff\hoch{4}}

\begin{document}

\hfill{KCL-TH-06-06}

\hfill{UUITP-10/06}

\hfill{HIP-2006-28/TH}

\hfill{USITP-06-03}

\hfill{hep-th/0607156}

\hfill{\today}

\vspace{20pt}

\begin{center}
{\Large{\bf On the covariance of the Dirac-Born-Infeld-Myers action}} \vspace{30pt}

\auth

\vspace{15pt}

\begin{itemize}
\item [$^1$] \kings \item [$^2$] \uu \item[$^3$] \hip
\item  [$^4$] \stock
\end{itemize}

\vspace{60pt}

{\bf Abstract}

\end{center}

A covariant version of the non-abelian Dirac-Born-Infeld-Myers action is presented. The non-abelian
degrees of freedom are incorporated by adjoining to the (bosonic) worldvolume of the brane a number
of anticommuting fermionic directions corresponding to boundary fermions in the string picture. The
proposed action treats these variables as classical but can be given a matrix interpretation if a
suitable quantisation prescription is adopted. After gauge-fixing and quantisation of the fermions,
the action is shown to be in agreement with the Myers action derived from T-duality. It is also
shown that the requirement of covariance in the above sense leads to a modified WZ term which also
agrees with the one proposed by Myers.

\pagebreak \tableofcontents \setcounter{page}{1}

%%%%%%%%%%%%%%%%%%%%%%%%%%%%%%%%%%%%%%%%%%%%%%%%%%%%%%%%%%%%%%%

\section{Introduction}

%%%%%%%%%%%%%%%%%%%%%%%%%%%%%%%%%%%%%%%%%%%%%%%%%%%%%%%%%%%%%%%%

An intriguing feature of string theory is the fact that one can have coincident D-branes. The
lowest-order action for such stacks of branes includes the Yang-Mills action for the non-abelian
gauge fields which arise when one takes the coincidence limit. Since the gauge part of the action
for a single D-brane is Born-Infeld one would expect that a non-abelian generalisation of this
action should be required in the case of coincident branes. Although there has been a lot of work
on this topic it is still not completely clear what this action is and how it should incorporate
invariance principles. It is the purpose of this note to propose such an action for both the
Dirac-Born-Infeld and the WZ terms which should be present. The derivation we give is strictly
speaking only valid in a certain approximation, which we explain below, but we shall argue that it
is not unreasonable to expect that it can be extended beyond this.

Many features of the bosonic terms in the non-abelian action are known. Some years ago, Tseytlin
\cite{Tseytlin:1997cs} put forward the proposal that the ordinary Born-Infeld action could be
generalised to the non-abelian case by using the same formula with an overall symmetrised trace.
Although this does not incorporate all the terms in the effective string action
\cite{Kitazawa:1987xj} it is nevertheless a well-defined object to work with. Subsequently,
starting from the Tseytlin action, Myers derived a non-abelian Dirac-Born-Infeld action by
demanding that lower-dimensional brane actions be consistent with T-duality \cite{Myers:1999ps}. He
also used T-duality to derive a non-abelian Wess-Zumino term and showed that this has the property
that higher degree RR forms can couple to a D$p$-brane giving rise to a dielectric effect. Similar
results were obtained from matrix theory \cite{Taylor:1999gq,Taylor:1999pr}. We shall show that
Myers's results can be understood from the point of view of invariance under diffeomorphisms of the
brane and gauge symmetries.

The Myers version of the DBI action was derived in the physical gauge where $(p+1)$ of the
coordinates of the target space are identified with those of the brane and the transverse
coordinates are taken to be the scalar fields. These are then promoted to matrices in the
non-abelian theory. This procedure clearly breaks diffeomorphism symmetry for the brane. Although
non-abelian gauge invariance is maintained through the use of covariantised pull-backs it is not at
all obvious that the Myers action is invariant with respect to gauge transformations of the
background gauge fields. This is also true for the Wess-Zumino term which involves the RR
potentials as well as the $B$ field. In fact, it is not clear what the non-abelian generalisation
of the modified field strength which appears in the action for a single D-brane should be. A
proposal for this was made in \cite{Howe:2005jz} in the supersymmetric context and this is the one
which we shall use here.

In this paper we shall derive non-abelian DBI and WZ actions in a formalism which is inspired by
the use of boundary fermions to describe non-abelian degrees of freedom in open string theory
\cite{Marcus:1986cm,Dorn:1996an,Kraus:2000nj,Berkovits:2002ag}. We shall work in the approximation
in which these fermions are taken to be classical variables. Mathematically this amounts to
extending the worldvolume of the brane by a number of fermionic directions and replacing the brane
embedding in the target spacetime by a generalised embedding defined as a map from the extended
worldvolume to the target space \cite{Howe:2005jz}. Although the formalism is not a fully-fledged
matrix formalism, the results we derive can be compared to those in the literature if we replace
the Poisson bracket (in the fermionic part of the space) by the matrix commutator and impose the
symmetrised trace prescription. This approach can be justified to some extent in the world-sheet
picture. In the papers cited above it is shown how quantisation of the fermions leads to the
fermions being replaced by gamma matrices, so that functions of them become matrices, and how
correlation functions of products of operators involving the fermions become the path-ordered trace
of products of matrices. Since we are concerned with operators at the same boundary point it is
natural to adopt the prescription that the path-ordered trace goes over to the symmetrised trace in
this case. Moreover, it is canonical practice to replace Poisson brackets by commutators in the
quantisation procedure.

The DBI action we propose has a very simple structure. Since the extended space is actually a
superspace it is natural to replace the Born-Infeld determinant with a superdeterminant. The matrix
in this superdeterminant is the sum of the pull-back of the target space metric and an abelian
two-form field strength modified by the pull-back of the $B$-field. The non-abelian gauge field
emerges from the expansion of the abelian gauge field in the fermionic coordinates. This action is
manifestly invariant under diffeomorphisms and gauge transformations and we show explicitly that it
reproduces the Myers DBI action in the physical gauge. Our WZ action looks very similar to the
Myers WZ action. We show that the couplings of the branes to scalar commutators can be motivated by
diffeomorphism invariance while the couplings to the higher rank RR forms are then required by RR
gauge symmetry. However, our formalism is not manifestly invariant and one has to work to prove
these results. Although the structure of our WZ action is very similar to Myers's there is a
difference in that his involves contractions of forms with the scalar commutator whereas ours
involves a similar contraction but in the fermionic directions. We show explicitly how the terms
collect together to gives the Myers result. Our conventions for differential forms are given in the
appendix.

%%%%%%%%%%%%%%%%%%%%%%%%%%%%%%%%%%%%%%%%%%%%%%%%%%%%%%%%%%%%%%%%

\section{The geometry of $\hM$ and the non-abelian gauge field}

%%%%%%%%%%%%%%%%%%%%%%%%%%%%%%%%%%%%%%%%%%%%%%%%%%%%%%%%%%%%%%%%%%

We shall be interested in a $p$-brane specified by an embedding $f:M\rightarrow \unM$ where $M$ is
the worldvolume of the brane and $\unM$ the target space, both spaces being bosonic. In order to
incorporate non-abelian degrees of freedom we extend the former to a superspace, $\hM$, the
coordinates of which we denote by $y^M=(x^m,\h^\m)$. The coordinates of the target space are
denoted by $x^{\unm}$. The embedding is replaced by a generalised embedding
$\hf:\hM\rightarrow\unM$. The space $\hM$ is equipped with an abelian gauge field $\cA$ such that
the modified field strength

\be
 K:=d\cA-\hf^* B\ ,
 \la{2.0}
\ee

where $B$ is the NS two-form potential on the target space, is invariant under gauge
transformations of both objects. It is furthermore assumed that $K_{\m\n}$ is non-singular.

Following the ideas developed in \cite{Howe:2005jz} we use the field $K$ to specify horizontal
subspaces in the tangent spaces of $\hM$. If $\o$ is a one-form on $\hM$ then we define its
horizontal component to be

\be
 \gho_m=\o_m- K_m{}^{\n}\o_{\n}\ ,
 \la{2.1}
\ee

where

\be
 K_m{}^{\m}:=K_{m\n} N^{\n\m}
 \la{2.2}
\ee

and

\be
 N^{\m\n}:=(K_{\m\n})^{-1}\ .
 \la{2.3}
\ee

We can view this as a change of basis if we also identify $\gho_{\m}=\o_{\m}$. For a vector $v$ we
have

\bea
 \hv^{m}&=&v^m\nn\w1
 \hv^{\m}&=&v^{\m} + v^m K_m{}^{\m}
 \la{2.4}
\eea

We define the horizontal component of $K$ itself to be $\cF$,

\be
 \cF_{mn}:=K_{mn}-K_{m\m} N^{\m\n} K_{\n n}\ .
 \la{2.5}
\ee

We note that $K$ has no mixed components in the hatted basis. It is straightforward to compute the
transformation properties of various objects under diffeomorphisms. In particular, we have

\bea
 \d K_{m}{}^{\m}&=& \hv^n(\cD_n K_m{}^{\m}-\cD_m K_n{}^{\m})+\hv^{\n}\del_{\n}K_m{}^{\m}+\cD_m
 \hv^{\m} - N^{\m\n}\del_{\n}\hv^n\cF_{nm}\nn\w1
 \d N^{\m\n}&=&\hv^m\cD_m N^{\m\n} + \hv^{\r}\del_{\r}N^{\m\n}-2N^{\r(\m}\del_{\r}\hv^{\n)}+2
 \hv^m N^{\r(\m}\del_{\r}K_m{}^{\n)}
 \la{2.6}
\eea

which implies that

\bea
 \d \gho_m &=&\hv^n \cD_n\gho_m + \hv^{\n}\del_{\n} \gho_m + \cD_m\hv^n \gho_n +
 N^{\m\n}\del_{\n}\hv^n \cF_{nm} \gho_{\m}\nn\w1
 \d\gho_{\m}&=&\hv^n\cD_n\gho_{\m}+\hv^{\n}\del_{\n}\gho_{\m}+\del_{\m}\hv^n\gho_n+
 \del_{\m}\hv^{\n}\gho_{\n}-\hv^n\del_{\m} K_n{}^{\n}\gho_{\n}\ .
 \la{2.7}
\eea

The derivative $\cD_m$ is defined by

\be
 \cD_m:=\del_m-K_m{}^{\m}\del_{\m}\ .
 \la{2.8}
\ee

The transformation rule of $\gho_m$ shows that this object is not preserved under general
diffeomorphisms, only those for which $\del_{\m} \hv^n=0$. Transformations which do not satisfy
this constraint are needed in order to reach the physical gauge where the generalised embedding has
the form

\be
 x^{\unm}=(x^m, x^{m'}(x,\h))\ .
 \la{2.9}
\ee

However, we do not want to make this gauge choice at this stage as the power of covariance would be
lost. Note that any object which has no mixed components in the hatted basis will transform
homogeneously under diffeomorphisms. This holds for $\cF_{mn}$.

We now turn to the emergence of a non-abelian gauge field from the abelian one we have introduced.
The requirement that $K_{\m\n}$ be non-singular will be satisfied for any background $B$ if
$(d\cA)_{\m\n}$ is non-singular. We can then use a vertical diffeomorphism to bring $\cA_{\m}$ to
the standard form

\be
 \cA_{\m}=\frac{1}{2}\h_{\m}\ .
 \la{2.10}
\ee

The transformations of $\cA$ are

\be
 \d\cA_M=\del_M a + v^N(\del_N \cA_M-\del_N\cA_N) + b_M
 \la{2.11}
\ee

where $a$ is the abelian gauge parameter and $b_M$ is the pull-back of the gauge parameter for
gauge transformations of the $B$ field. As stated above we can use $v^{\m}$ to go to the standard
gauge for $\cA_{\m}$. The residual vertical diffeomorphisms are then given by

\be
 v^{\m}=-\d^{\m\n}(\del_{\n} a + b_{\n}-v^n \del_{\n}\cA_n)
 \la{2.12}
\ee

We shall denote $\cA_m$ in the standard gauge by $A_m$; it transforms as

\be
 \d A_m=\del_m a + (A_m,a) + \tb_m + v^n F_{nm}\ .
 \la{2.13}
\ee

where the Poisson bracket $(,)$ is defined by

\be
 (f,g):=\d^{\m\n} \del_{\m} f\del_{\n}g\ .
 \la{2.14}
\ee

$F_{mn}:=\del_m A_n-\del_n A_m+ (A_m,A_n)$ is the non-abelian field strength tensor, and  $\tb_m$
denotes the covariant pull-back of $b$ with respect to the Yang-Mills derivative. This is given by

\be
 \tb_m:=D_m x^{\unm}b_{\unm}=(\del_m x^{\unm}+A_m{}^{\m} \del_{\m} x^{\unm}) b_{\unm}\ ,
 \la{2.15}
\ee

where

\be
 A_m{}^{\m}:=\d^{\m\n} \del_{\n} \cA_m\ .
 \la{2.16}
\ee

The relation between the non-abelian field strength tensor and $\cF$, in the standard gauge, is
given by

\be
 \cF_{mn}=F_{mn}-\tB_{m n}- \tB_{m \m} N^{\m\n}\tB_{\n n}\ .
 \la{2.17}
\ee

This formula may be taken as the definition of the appropriately modified non-abelian field
strength tensor in the presence of a $B$ field.

For later use we note the relation between the hatted and tilded bases, valid in the standard
gauge. The fermionic components of a one-form $\o$ are the same while

\be
 \gho_m=\wt{\o}_m + \tB_{m\m} N^{\m\n} \o_{\n}\ .
 \la{2.18}
\ee

%%%%%%%%%%%%%%%%%%%%%%%%%%%%%%%%%%%%%%%%%%%%%%%%%%%%%%%%%%%%%%%

\section{The DBIM action}

%%%%%%%%%%%%%%%%%%%%%%%%%%%%%%%%%%%%%%%%%%%%%%%%%%%%%%%%%%%%%

In this section we present the Lagrangian for the DBIM action in the presence of the additional
fermionic variables. We set

\be
 L_{MN}:= g_{MN} + K_{MN}
 \la{3.1}
\ee

where $g_{MN}$ denotes the pull-back of the target-space metric to $\hM$. The Lagrangian is then
simply

\be
 \cL=\sqrt{-{\sdet}\, L_{MN}}\ .
 \la{3.2}
\ee

This Lagrangian obviously transforms as a density under diffeomorphisms of $\hM$ and is manifestly
invariant under gauge transformations of both $\cA$ and $B$. We shall now show that it coincides
with the Myers action \cite{Myers:1999ps} in the physical gauge provided that we interpret
functions of $\h$ as matrices, replace Poisson brackets by commutators and replace integration over
the fermionic variables with the symmetrised trace over all matrix factors.

The superdeterminant is

\be
 {\sdet}\,{L_{MN}}= \det (L_{mn}-L_{m\m}L^{\m\n} L_{\n n})(\det L_{\m\n})^{-1}\ ,
 \la{3.3}
\ee

where $L^{\m\n}:=(L_{\m\n})^{-1}$. If we introduce $E_{MN}:=g_{MN}-B_{MN}$, then

\be
 L_{mn}=E_{mn} + (dA)_{mn}\
 \la{3.4}
\ee

while

\be
 L_{\m\n}=\d_{\m\n}+E_{\m\n}
 \la{3.5}
\ee

and

\be
 L_{m\m}=E_{m\m}-A_{m\m}
 \la{3.6}
\ee

in the standard gauge, $A_{\m}={1\over2}\h_\m$. We remind the reader that $A_{m\m}=\del_{\m}
A_m$.We therefore have

\be
 L_{mn}-L_{m\m}L^{\m\n} L_{\n
 n}=E_{mn}+(dA)_{mn}-(E_{m\m}-A_{m\m})(\d_{\m\n}+E_{\m\n})^{-1}(E_{\n
 n}+A_{n\n})\ .
 \la{3.7}
\ee

We want to express this in terms of $F_{mn}$ and replace the ordinary pull-back on $m$ indices by
the covariant one defined by $D_m=\del_m + A_m{}^{\m}\del_{\m}$. After a straightforward piece of
algebra one indeed finds that

\be
 L_{mn}-L_{m\m}L^{\m\n} L_{\n n}=\tE_{mn}+F_{mn}-\tE_{m\m} (\d_{\m\n}+E_{\m\n})^{-1}
 \tE_{\n n}\ ,
 \la{3.8}
\ee

where

\bea
 \tE_{mn}&=& D_m x^{\unm} D_n x^{\unn} E_{\unm\unn}\nn\w1
 \tE_{m\n}&=& D_m x^{\unm} \del_\n x^{\unn} E_{\unm\unn}\ .
 \la{3.9}
\eea

We shall now compute the second term on the RHS of \eq{3.8} in the physical gauge,
$x^{\unm}=(x^m,x^{m'}(x,\h))$. In this gauge

\bea
 \tE_{m\n}&=& D_m x^{\unm}\del_{\n} x^{\unn} E_{\unm\unn}\nn\w1
 &=& D_m x^{\unm}\del_{\n} x^{n'} E_{\unm n'}\nn\w1
 &=& \del_{\n} x^{n'} (E_{m n'}+A_m{}^{\m}\del_{\m} x^{m'} E_{m'n'})\nn\w1
 &=& \tE_{m n'} \del_{\n} x^{n'}\ ,
 \la{3.10}
 \eea

while

\be
 E_{\m\n}=\del_{\m} x^{m'} \del_{\n} x^{n'} E_{m'n'}\ .
 \la{3.11}
\ee

We therefore have

\be
 \tE_{m\m} (\d_{\m\n}+E_{\m\n})^{-1}\tE_{\n n}=
 \tE_{m p'} \del_{\m}x^{p'}(\d_{\m\n}+E_{\m\n})^{-1}\del_{\n} x^{q'}
 \tE_{q' n}\ .
 \la{3.12}
\ee

Expanding out the inverse we find

\be
 \del_{\m}x^{p'}(\d_{\m\n}+E_{\m\n})^{-1}\del_{\n} x^{q'}=
 M^{p'q'}- M^{p'r'} E_{r's'} M^{s'q'} + \ldots
 \la{3.13}
\ee

where

\be
 M^{m'n'}:= \d^{\m\n}\del_{\m} x^{m'} \del_{\n} x^{n'}=(x^{m'},x^{n'})\ .
 \la{3.14}
\ee

The series of terms is easily summed; the final result is

\be
 L_{mn}-L_{m\m}L^{\m\n} L_{\n n}=F_{mn} + \tE_{mn}+\tE_{m p'} \left(
 (Q^{-1}-1) E^{-1}\right)^{p'q'} \tE_{q' n}
 \la{3.15}
\ee

where

\be
 Q^{m'}{}_{n'}=\d^{m'}{}_{n'} + M^{m' p'} E_{p' n'}\ ,
 \la{3.16}
\ee

and where $E^{-1}$ denotes the inverse of $E_{m'n'}$.

In order to complete the picture we need to compute $\det L_{\m\n}$ and show that it equals
$\det^{-1} Q$ in the physical gauge. This is a straightforward exercise using the $\exp \tr \ln$
formula for the determinant. For example, one has

\bea \d^{\m\n} E_{\m\n}&=&\d^{\m\n}\del_{\m} x^{m'} \del_{\n} x^{n'} E_{m'n'}\nn\w1
 &=&M^{m'n'}E_{m'n'}\nn\w1
 &=&-\tr (ME)
 \la{3.17}
\eea

in the physical gauge. One therefore obtains

\be
 \det L_{\m\n}=\exp(-\tr\ln(1+ME))=(\det Q)^{-1}
 \la{3.18}
\ee

as required. So we indeed find that

\be
 \sqrt{-{\sdet}\,L_{MN}}=\sqrt{-\det\,(\tE_{mn} + F_{mn}+\tE_{m
 p'}[(Q^{-1}-1)E^{-1}]^{p'q'}\tE_{q' n}) \det\,Q}\ .
 \la{3.19}
\ee

It is quite remarkable that this expression agrees precisely with Myers's result
\cite{Myers:1999ps} provided that one interprets it in the way  we have suggested.

%%%%%%%%%%%%%%%%%%%%%%%%%%%%%%%%%%%%%%%%%%%%%%%%%%%%%%%%%%%%%%%%%%%%%%%%%%%%

\section{The Wess-Zumino term}

%%%%%%%%%%%%%%%%%%%%%%%%%%%%%%%%%%%%%%%%%%%%%%%%%%%%%%%%%%%%%%%%%%%%%%%%%

In \cite{Myers:1999ps} Myers gives an expression for the WZ term for a D$p$-brane modified to the
non-abelian case. The most remarkable feature of this term is that $p$-branes can couple to
higher-degree RR potential forms. The mechanism for this is that one can lower the degree of a
pulled-back form by two by contracting a pair of transverse indices with $M^{m'n'}$, which is the
commutator in Myers. In this way, for example, a five-form RR field can give rise to a three-form
and thus couple to a D2-brane. However, it is far from obvious that Myers's WZ term is
gauge-invariant or invariant under diffeomorphisms of the brane. In this section we construct a WZ
term in our model which has these properties, although not manifestly. Its structure is very
similar to the Myers WZ term, although the match up of the terms is not quite straightforward.

The Wess-Zumino term for a D$p$-brane is

\be
 \cL_{WZ}= \sqrt{N}\left[\exp(-\frac{1}{2}i_N) e^{\cF} \sum
 \hC\right]_{p+1,0}\ ,
 \la{4.1}
\ee

where $N:=\det\, N^{\m\n}$. The subscript $(p+1,0)$ indicates that the $(p+1,0)$-form component is
to be projected out, a $(p,q)$-form being a $(p+q)$-form on $\hM$ with even degree $p$ and odd
degree $q$ in the hatted basis. The $RR$ potentials $\hC$ are pulled back to $\hM$ with the hatted
pull-back, e.g.

\be
 \hC_{m \n}=\cD_m x^{\unm} \del_{\n} x^{\unn} C_{\unm\unn}\ ,
 \la{4.2}
\ee

where

\be
 \cD_m=\del_m -K_m{}^{\m}\del_{\m}.
 \la{4.3}
\ee

The point of using this pull-back rather  than the one defined with the straightforward gauge
covariant derivative is that it is invariant with respect to the abelian gauge transformations of
both $\cA$ and $B$. The operation $i_N$ appearing in \eq{4.1} denotes the contraction of a form
with $N^{\m\n}$. Again this is invariant under gauge transformations of $\cA$ and $B$. However, the
full expression is neither manifestly covariant under diffeomorphisms of $\hM$ nor under gauge
transformations of the RR fields.

First suppose that the WZ form for a D$p$-brane, divided by $\sqrt{N}$,  transforms in a regular
fashion, i.e. without a term involving $\del_{\m} \hv^m$, and let $L$ be the dual of this form,
then we can take the Lagrangian, regarded as a function, to be $L_{WZ}=\sqrt{N} L$, and we have

\be
 \d L=(\hv^m \cD_m  + \hv^{\m} \del_{\m}) L + \cD_m \hv^m L
 \la{4.4}
\ee

while

\be
 \d \sqrt{N}=(\hv^m \cD_m  + \hv^{\m} \del_{\m}) \sqrt{N} +
 \sqrt{N} K_{\m\n}(\hv^m
 N^{\m\r}\del_{\r}K_m{}^{\n}-N^{\m\r} \del_{\r}\hv^{\n})\ .
 \la{4.5}
\ee

Combining these two results it is easy to see that

\be
 \d (\sqrt{N}L)=(-1)^M \del_M (v^M (\sqrt{N}L))\ .
 \la{4.6}
\ee

We therefore see that if $L$ transforms regularly then $L_{WZ}$ will transform in the desired
fashion under diffeomorphisms of $\hM$. The transformations of the pulled-back RR forms do include
irregular terms and so the problem is to show that these all cancel between the different terms in
the action involving the same RR field but different powers of $i_N$. In fact, we can use this to
argue that these terms must be present. As a simple example, consider the $C_3$ terms in the action
for a D2-brane. If we focus on only the irregular parts of the transformation of $C_3$ we have

\be
 \d \hC_{mnp} \sim 3 N^{\m\n} \del_{\n} \hv^q \cF_{q[m} \hC_{|\m|np]}
 \la{4.7}
\ee

Clearly we need to add something to the Lagrangian proportional to $N$ to cancel this variation.
The obvious expression to try is $i_N \hC\cF$, which involves the part of $C$ with two fermionic
indices. Since $\cF$ has no fermionic indices, this term involves $\hC_{\m\n p}$. The irregular
terms in the transformation of this field are

\be
 \d \hC_{\m\n p}\sim 2\del_{(\m} \hv^q \hC_{|q|\n )p}+ N^{\r\s} \del_{\r}\hv^q \cF_{qp}
 \hC_{\m\n\s}\ .
 \la{4.7.1}
\ee

The terms of interest in the transformation of $i_N \hC\cF$ are therefore

\bea
 \d(N^{\m\n} \hC_{\m\n [m} \cF_{np]})&\sim &
 N^{\m\n} N^{\r\s} \del_{\r} \hv^q \hC_{\m\n\s}\cF_{q[m}\cF_{np]} \nn\w1
 &\phantom{~}&  + 2 N^{\m\n} \del_{\m}\hv^q \hC_{q\n[m }\cF_{np]}\ .
 \la{4.8}
\eea

The first line on the RHS vanishes as it involves $\cF\wedge\cF$ in three dimensions, while the
second is equal to +2 multiplied by the RHS of \eq{4.7}. Expanding out the exponential in \eq{4.1}
we see that precisely the right coefficient is generated in order for these two terms to cancel.
The remaining terms can easily be seen to be the regular terms in the transformation of the part of
the D2-brane action which involves $C_3$.

This line of reasoning can easily be extended to the general case. For a given brane, a given RR
field will appear in a sequence of terms with increasing powers of $\cF$ and $i_N$. It is not
difficult to verify that the irregular variations in a given term cancel against those coming from
the two adjacent terms. We can therefore conclude that the WZ term is invariant up to a total
derivative under abelian gauge transformations of $\cA$ and $B$ and diffeomorphisms of $\hM$. Since
the non-abelian gauge transformations arise from the vertical diffeomorphisms combined with the
abelian gauge transformations of $\cA$ in the standard gauge we are therefore assured that the WZ
term will be invariant under these.

The above considerations indicate the need for the $i_N$ terms but do not mix RR forms of different
rank. The full structure is required by demanding gauge invariance for the background RR fields.
Since the WZ Lagrangian is a sum of standard WZ terms of different rank one might think that this
is obvious but closer inspection shows that it is not, because the $i_N$ operation does not commute
with exterior differentiation.

The proof of gauge invariance is not difficult however. We shall focus on the IIA case for
simplicity. The gauge transformations of a $(p+1)$-form potential is

\be
 \d C_{p+1}=d \L_{p} - \L_{p-2} H
 \la{4.9}
\ee

where $H$ is the NS three-form, and $\L$ is used to denote the gauge parameters. Pulled back to
$\hM$, $H$ becomes $-dK$. So the gauge transformation of the WZ form is

\be
 \d \cL_{WZ}=\sqrt{N}\left[e^{-\frac{1}{2}i_N}  \sum (d(\ghL e^{\cF}) + d N^{-1}\ghL e^{\cF})\right]_{p+1,0}\ ,
 \la{4.10}
\ee

where

\be
 N^{-1}:=K-\cF
 \la{4.11}
\ee

Now $\cF$ is horizontal and is not acted on by $i_N$. It therefore plays no essential r\^{o}le in
the proof that this expression gives rise to a total derivative. From this point of view it may as
well be absorbed into the parameters $\L$; that is, we regard $\L e^{\cF}$ as the redefined $\L$.
Each $\L$ appears in pairs of terms which have the same degree. A general pair of terms in
\eq{4.10} has the form

\be
 (-\frac{1}{2})^n\sqrt{N}\left( i_N^n (d\ghL)_{p+1,2n}-\frac{1}{2n+2}i_N^{n+1}
 (dN^{-1}\ghL)_{p+1,2n+2}\right)\ .
 \la{4.12}
\ee

This gives the contribution of a $(2n+p)$-form $\ghL$. We shall now show that this expression can
indeed be written as a total derivative. The above formula suggests that it should be something
like $d(\sqrt{N}i_N^n \ghL)$ and we shall see that this is not far off the mark.

Let us introduce the notation $N^{\m_1\ldots \m_{2n}}$ to denote the symmetrised product of $n$
$N$s. The first term in \eq{4.12}, ignoring the overall factor and the square root, is

\bea
 &&N^{\m_1\ldots \m_{2n}}(d\ghL)_{m_1\ldots m_{p+1}\m_1\ldots \m_{2n}}\nn\w2
 &=& N^{\m_1\ldots \m_{2n}}\Big( (p+1)\cD_{m_1}\ghL_{m_2\ldots m_{p+1}\m_1\ldots \m_{2n}}
   -2n\del_{\m_1}\ghL_{m_1\ldots m_{p+1}\m_2\ldots \m_{2n}}\nn\w2
 &\phantom{=} &-p(p+1) \cD_{m_1} K_{m_2}{}^{\r} \ghL_{m_3\ldots m_{p+1} \m_2\ldots \m_{2n}\r}
 -2n(p+1) \del_{\m_1}K_{m_1}{}^{\r} \ghL_{m_2\ldots m_{p+1} \m_2\ldots \m_{2n}\r}\Big)\ ,\nn\w1
 &&
 \la{4.14}
\eea

where antisymmetrisation over the free even indices is understood here and in the rest of the
proof. The terms involving the derivatives of $K$ arise because we are working in the horizontal
lift basis. The second term in \eq{4.12}, again omitting the numerical factors and the square root,
is

\bea
 &&N^{\m_1\ldots \m_{2n+2}}(dN^{-1})_{[m_1 m_2 m_3} \ghL_{m_4\ldots m_{p+1}\m_1\ldots \m_{2n+2}]}\nn\w1
 &=&N^{\m_1\ldots \m_{2n+2}}\Big(\frac{(2n+2)p(p+1)}{2}
 (dN^{-1})_{m_1 m_2 \m_1} \ghL_{m_3\ldots m_{p+1}\m_2\ldots\m_{2n+2}}\nn\w1
 &\phantom{=}&\phantom{N^{\m_1\ldots \m_{2n+2}}}
 +\frac{(2n+2)!(p+1)}{2(2n)!}(dN^{-1})_{m_1 \m_1\m_2} \ghL_{m_2\ldots m_{p+1}\m_3\ldots \m_{2n+2}}\nn\w1
 &\phantom{=}&\phantom{N^{\m_1\ldots \m_{2n+2}}}
 +\frac{(2n+2)!}{3!(2n-1)!}(dN^{-1})_{\m_1\m_2\m_3}\ghL_{m_1\ldots m_{p+1}\m_4\ldots \m_{2n+2}}\Big)\ .
 \la{4.15}
\eea

The exterior derivative of $N^{-1}$ has the following non-trivial components

\bea
 (dN^{-1})_{m_1m_2\m}&=&2\cD_{[m_1} K_{m_2]}{}^{\r} K_{\r \m}\nn\w1
 (dN^{-1})_{m_1 \m_2\m_3}&=& \cD_{m_1}K_{\m_2\m_3}-2\del_{(\m_2}K_{|m_1}{}^{\r}K_{\r|\m_3)}\nn\w1
 (dN^{-1})_{\m_1\m_2\m_3}&=& 3\del_{(\m_1} K_{\m_2\m_3)}\ .
 \la{4.16}
\eea

We can group together like terms from the two original terms in \eq{4.12} taking into account the
relative numerical factor. It is easy to see that the terms involving $\cD_m K_n{}^{\r}$ cancel.
Reinstating the square root of $\det N$ we see that the other terms with the derivative $\cD_m$ can
be written in the form

\be
 (p+1)\cD_{m_1} (\sqrt{N} N^{\m_1\ldots \m_{2n}}\ghL_{m_2\ldots m_{p+1}\m_1\ldots \m_{2n}})\ .
 \la{4.17}
\ee

Next we consider the terms involving $\del_{\m} K_m{}^{\r}$. The term of this sort in \eq{4.14} is
cancelled by one of the terms in \eq{4.15} and we are left with

\be
 (p+1)\sqrt{N}N^{\m_1\ldots \m_{2n}}\del_{\r} K_{m_1}{}^{\r}\ghL_{m_2\ldots m_{p+1}\m_1\ldots \m_{2n}}
 \la{4.18}
\ee

The two expressions \eq{4.17} and \eq{4.18} combine to give

\be
 (p+1)\del_{m_1}(\sqrt{N} (i_N^n \ghL)_{m_2\ldots m_{p+1}}) +
 (p+1)\del_{\m} (\sqrt{N} K_{m_1}{}^{\m} (i_N^n \ghL)_{m_2\ldots m_{p+1}})\ .
 \la{4.19}
\ee

Finally, we consider the terms involving the fermionic derivative of $\ghL$ and $N$. After a short
piece of algebra, and using the fact that $N^{\m\n}$ is the inverse of $K_{\m\n}$, one verifies
that these give

\be
 -2n\del_{\m_1}(\sqrt{N} N^{\m_1\ldots \m_{2n}} \ghL_{m_1\ldots m_{p+1}\m_2\ldots \m_{2n} })\ .
 \la{4.20}
\ee

To conclude the proof,  we can contract the above expressions, \eq{4.19} and \eq{4.20}, with
$\e^{m_1\ldots m_{p+1}}$ to obtain total derivatives. We therefore conclude that the WZ term is
indeed invariant, up to a divergence in $\hM$, under gauge transformations of the background RR
potentials.

%%%%%%%%%%%%%%%%%%%%%%%%%%%%%%%%%%%%%%%%%%%%%%%%%%%%%%%%%%%%%%%%%%%%%%%%%%%%%%%%%%

\section{Comparison of WZ terms}

%%%%%%%%%%%%%%%%%%%%%%%%%%%%%%%%%%%%%%%%%%%%%%%%%%%%%%%%%%%%%%%%%%%%%%%%%%%%%%%%%%%%

The structure of our WZ term is clearly very similar to that of Myers's, but the way in which the
various terms match up is not obvious since our inner product operation $i_N$ involves the extra
fermionic directions while Myers's involves the directions transverse to the brane in spacetime. In
this section we shall prove that the two expressions agree precisely in the static gauge, given
that we interpret our results as before.

The WZ term for a $p$-brane is

\be
 \cL_{WZ}=\sqrt{N}\left[e^{-\half i_N}e^{\cF}\sum \hC\right]_{p+1,0}
 \la{5.1}
\ee

$\cF$ is a $(2,0)$-form which can be written, in the standard gauge \eq{2.17},
as\footnote{Throughout this section forms are written in the hatted basis
$(e^m=dx^m,\he^{\m}=d\h^{\m} + dx^m K_m{}^{\m})$. Tildes refer to the components, e.g.
$\tB_{1,1}=\he^\m e^m \tB_{m\m}$.}

\be
 \cF_{2,0}=F_{2,0}-\tB_{2,0} - b_{2,0}\ ,
 \la{5.2}
\ee

where

\be
 b_{mn}:=\tB_{m\m} N^{\m\n}\tB_{\n n}\ .
 \la{5.3}
\ee

The first two terms in $\cF$ are very similar in our expression and in Myers's; since they are not
seen by the $i_N$ operation they can be absorbed into the RR potentials. With this understanding we
can write the contribution of a $(p+1+2n)$-form potential to the WZ term as

\bea
 \cL_{WZ} &\sim & \sqrt{N}\left[e^{-\half i_N}e^{-b_{2,0}}\hC_{p+1+2n}\right]_{p+1,0}\nn\w2
 &=& \sqrt{N}\left[e^{-\half
 i_N}\left(\hC_{p+1,2n}-b_{2,0}\hC_{p-1,2n+2}+\ldots\right)\right]_{p+1,0}\ .
 \la{5.4}
\eea

We shall prove the equivalence between this expression and the corresponding Myers term in steps.
First we show that, in the standard gauge,

\be
 \left[e^{-\half i_N}e^{-b_{2,0}}\hC_{p+1+2n}\right]_{p+1,0}=
 \left[e^{-\half i_N}e^{-\tB_{1,1}}\tC_{p+1+2n}\right]_{p+1,0}\ ,
 \la{5.5}
\ee

where  $\tB_{1,1}$ has components $\tB_{m\m}$. In order to prove this result we shall need to
express $\hC$ in terms of $\tC$, the Yang-Mills pulled-back potential. It is straightforward to see
that

\be
 \hC_{p+1,2n}=\sum_{q=0}^{q=p+1} \frac{1}{q!} (\tB_1{}^1)^q \tC_{p+1-q,2n+q}\ ,
 \la{5.7}
\ee

where $\tB_1{}^1$ denotes the vector-valued one-form with components $\tB_{m \n} N^{\n
\m}:=\tB_m{}^{\m}$.

We begin by looking at the terms with no $b$-field contributions. The term with $q$ powers of $\tB$
on the LHS of \eq{5.5} is

\be
 (-\half)^n\, \frac{ i_N^n}{q! n!} (\tB_1{}^1)^q \tC_{p+1-q,2n+q}\ ,
 \la{5.9}
\ee

whereas the RHS contribution comes from the term

\be
 (-\half)^{n+q}\, \frac{(-1)^q i_N^{n+q}}{q!(n+q)!} \tB_{1,1}^q \tC_{p+1-q,2n+q}\
 \la{5.10}
\ee

Now

\bea
 [\tB_{1,1}^q \tC_{p+1-q,2n+q}]_{m_1\ldots m_{p+1},\m_1\ldots \m_{2n}\n_1\ldots \n_q\r_1\ldots \r_q}
 &=&\e_q(-1)^{q(p+1-q)}\,\frac{(p+1)!(2n+2q)!}{(p+1-q)!(2n+q)!}\,\xz\nn\w2
 &\phantom{=}& \xz \,\tB_{[m_1(\n_1}\ldots
 \tB_{m_q\n_q}\tC_{m_{q+1}\ldots m_{p+1}] \m_1\ldots \m_{2n}\r_1\ldots \r_q)}\ \nn\w1
 &&
 \la{5.11}
\eea

where

\be
 \e_q:=(-1)^{\half q(q-1)}\ .
 \la{5.8}
\ee

In \eq{5.10} the expression in \eq{5.11} is to be contracted with $(n+q)$ powers of $N$. We can
compute this by switching the symmetrisation brackets on the fermionic indices to $N^{n+q}$. To
find the term with no $b$ we then have to isolate the term in this expression of the form
$N^{\m_1\ldots \m_{2n}}N^{\n_1\r_1}\ldots N^{\n_q\r_q}$. The result is

\be
 \e_q(-\half)^n\, \frac{i_N^n (p+1)!}{n!q!(p+1-q)!}\tB_{[m_1}{}^{\n_1}\ldots
 \tB_{m_q}{}^{\n_q} \tC_{m_{q+1}\ldots m_{p+1}]\m_1\ldots \m_{2n}\n_1\ldots \n_q}\ ,
 \la{5.12}
\ee

and this is exactly equal to the component form of \eq{5.9}.

Now consider the terms with $q$ powers of $\tB$ including $b^r$, where $q\geq 2r$. The LHS of
\eq{5.5} gives

\be
 (-\half)^{n+r}\, \frac{1}{(q-2r)!}\frac{i_N^{n+r}}{(n+r)!}\frac{(-1)^r b^r}{r!}
 \, (\tB_1{}^1)^{q-2r} \tC_{p+1-q,2n+q}\ ,
 \la{5.13}
\ee

We have to compare this with the $b^r$ term in \eq{5.10}. Using the fact that $i_N (\tB_{1,1})^2=4
b$, we find that this contribution to \eq{5.10} is

\be
 4^r n_{r,q,n}(-\half)^{n+q} \frac{(-1)^q}{q!}\frac{i_N^{n+q-r}}{(n+q)!}b^r\, \tB_{1,1}^{q-2r}
 \tC_{p+1-q,2n+q}\ ,
 \la{5.14}
\ee

where it is understood that the remaining contractions do not give any more factors of $b$. The
combinatoric factor $n_{r,q,n}$ is given by

\be
 n_{r,q,n}=\frac{q!(n+q)!}{2^r r!(q-2r)!(n+q-r)!}\ .
 \la{5.15}
\ee

Using this in \eq{5.14} and the previous result for no $b$ terms one can easily verify that the LHS
and RHS terms with $b^r$ in \eq{5.5} are indeed equal.

The second step is to show that

\be
 \sqrt{N}\left[e^{-\half i_N}\o_{q,2n}\right]_{q,0}=
 \left[e^{-\half i_\d}e^{-B_{0,2}}\o_{q,2n}\right]_{q,0}
 \la{5.16}
\ee

where $i_\d$ means contraction with $\d^{\m\n}$ instead of $N^{\m\n}$. The even part of the form
$\o$ does not play an essential r\^{o}le here, so we can take $q=0$. In this case the LHS of
\eq{5.16} is

\be
 \sqrt{N}\frac{(-\half)^n }{n!} i_N^n \o=\sqrt{N}(-\half)^n \frac{1}{n!}\sum_{\{k\}}
 \o(B^{k_1},\ldots ,B^{k_n})\ ,
 \la{5.17}
\ee

where, on the RHS, $\o$ is regarded as a symmetric $n$-linear map of symmetric matrices,
$B(=B_{0,2})$ is regarded as a symmetric matrix, $(k_1,\ldots k_n)$ is an $n$-tuple of non-negative
integers, and the sum is over all such $n$-tuples. $B^{k_i}$ denotes the $k_i$th power of $B$ as a
matrix, with $(B^0)^{\m\n}=\d^{\m\n}$. On the other hand, the RHS of \eq{5.16} is

\be
 \sum_{m}(-1)^m (-\half)^{n+m} \frac{i_\d^{n+m}(B_{0,2}^m\o)}{(n+m)!m!}
 \ .
 \la{5.18}
\ee

Now

\be
 \i_{\d}^{n+m} (B_{0,2}^m \o)=\sum_{\{k\}}^m \frac{2^k(n+m)!m!}{n!(m-k)!(m-k)!}\o(B^{k_1},\ldots
 ,B^{k_n})
 \i_{\d}^{m-k} B_{0,2}^{m-k}\ ,
 \la{5.19}
\ee

where $k=\sum_{i=1}^n k_i$, and where the sum runs over all $n$-tuples such that $k\leq m$. Thus
the RHS of \eq{5.16} is

\be
  \sum_{m,\{k\}}(-\half)^n \frac{1}{2^{m-k}[(m-k)!]^2}\o(B^{k_1},\ldots ,B^{k_n})
  \i_{\d}^{m-k} B_{0,2}^{m-k}\ .
 \la{5.20}
\ee

If we set $p=m-k$ we find that this is equal to

\be
 \left(\sum_p \frac{1}{2^p}\frac{i_\d^p B_{0,2}^p}{(p!)^2}\right)\sum_{\{k\}}
 (-\half)^n\frac{1}{n!}\o(B^{k_1},\ldots ,B^{k_n})
 =\left[e^{-\half i_\d}e^{-B_{0,2}}\right]_{0,0}\sum_{\{k\}}
 (-\half)^n\frac{1}{n!}\o(B^{k_1},\ldots ,B^{k_n})\ .
 \la{5.21}
\ee

It is not difficult to show that the factor in square brackets on the RHS of this equation is the
square root of the determinant,

 \be
 \sqrt{N}=\left[e^{-\half i_\d}e^{-\tB_{0,2}}\right]_{0,0}\ .
 \la{5.22}
\ee

The use of this in \eq{5.21}  establishes \eq{5.16}.\footnote{There is no difference between
$\tB_{0,2}$ and $B_{0,2}$}

Finally, in the physical gauge,

\be
 (i_{\d}^n) \o_{p,2n}=(-i_M)^n \o_{p,2n}\
 \la{5.23}
\ee

for any pulled-back form $\o_{p,2n}$, where, on the right,  $\o_{p,2n}$ is a $(p+2n)$-form on the
brane  tangential degree $p$ and normal degree $2n$. This is easy to see: suppose we have a
$(0,2)$-form, $\o_{0,2}$, then

\bea
 i_{\d} \o_{0,2}&=&\d^{\m\n} \o_{\m\n}\nn\w1
 &=&\d^{\m\n} \del_{\m} x^{m'}\del_{\n} x^{n'} \o_{m' n'}\nn\w1
 &=&M^{m'n'}\o_{m'n'}=-i_M \o \ .
 \la{5.24}
\eea

The proof can easily be extended to the general case. Combining all of these steps, and reinstating
$\cF$ we finally arrive at

\bea
 \sqrt{N}\left[e^{-\half i_N}e^{\cF}\sum \hC\right]_{p+1,0}&=&
 \sqrt{N}\left[e^{-\half i_N}e^{F-\tB_{2,0}}e^{-\tB_{1,1}}\sum \tC\right]_{p+1,0}\nn\w2
 &=&
 \textbf{}\left[e^{-\half i_\d}e^{F-\tB_{2,0}-\tB_{1,1}-\tB_{0,2}}\sum \tC\right]_{p+1,0}\nn\w2
 &=&
 \left[e^{\half i_M}e^{F-\tB}\sum \tC\right]_{p+1,0}
 \la{5.25}
\eea

The final expression is the Myers WZ term which is only defined in the physical gauge. In the Myers
term $\tB$ is the gauge-covariant pull-back of $B$. The equality between our WZ term and Myers's is
to be interpreted in the same manner as for the DBIM part of the action.

%%%%%%%%%%%%%%%%%%%%%%%%%%%%%%%%%%%%%%%%%%%%%%%%%%%%%%%%%%%%%%

\section{Discussion}

%%%%%%%%%%%%%%%%%%%%%%%%%%%%%%%%%%%%%%%%%%%%%%%%%%%%%%%%%%%%%%%%%

In this article we have argued that the action proposed by Myers for coincident D-branes in the
physical gauge can be derived from a completely covariant formalism which makes use of boundary
fermions. In order to make the final step from our results to Myers's we have to replace functions
of $\h$ by matrices and replace integration over the fermionic variables by the symmetrised trace
over matrices. This is quite natural as the boundary fermions have to be quantised when they are
used to describe the Chan-Paton factors for the open string. In the standard gauge canonical
quantisation leads to $\h^{\m}$ being replaced by $\c^{\m}$, so that functions of fermions
naturally give rise to matrices. The path integral over the boundary fermions is designed to
reproduce the path-ordered trace; if the integrand is local as a function of the parameter
specifying the boundary of the string it is natural to interpret this as the symmetrised trace. It
would nevertheless be preferable to justify this prescription in more detail from the worldsheet
point of view, a topic we hope to report on in the future.

The main feature of our result is that the Myers action can be viewed as a gauge-fixed version of
an action which is covariant under all the local symmetries. Although one would have expected this
to be the case, it was not at all clear how the Myers action could be made compatible with these.
Since the Myers action is defined in the physical gauge diffeomorphism invariance could not have
been present, but invariance under gauge transformations of the background potentials was also
rather obscure.

In \cite{Howe:2005jz} we used the boundary fermion formalism in the context of the open
Green-Schwarz string to derive the equations of motion for a set of coincident supersymmetric
branes in the classical limit, i.e. taking the fermions to be classical. Given the results of the
present paper these equations should be equivalent to those one would derive from a
supersymmetrisation of Myers's action. However, it is not easy to make the comparison because the
Myers formalism is most suited to the action whereas the formalism of \cite{Howe:2005jz} leads
naturally to equations of motion. It is not so easy to relate the two in such a complicated theory.

In \cite{Howe:2005jz} we used a supersymmetric extension of the formalism employed here, in the
sense that the brane and the target space were taken to be superspaces, and in that context we also
developed a manifestly covariant formalism. We then attempted to construct a covariant action form
by generalising the construction of brane actions \cite{Bandos:1995dw,Howe:1998ts} in the
superembedding formalism \cite{Bandos:1995zw,Howe:1996mx}. Although the details of this were not
fully worked out the results derived in the present paper suggest that they could be. If so, this
could be a much quicker and more transparent way of establishing covariance. A related topic,
currently under investigation, is the construction of a kappa-symmetric Green-Schwarz action for
coincident D-branes. It would also be interesting to compare the current approach with that of
\cite{Drummond:2002kg} where only a single kappa-symmetry is employed.

In conclusion, the current work suggests that the boundary fermion formalism could be very useful
for discussing non-commutative aspects of coincident D-branes. Clearly further work needs to be
done to put this approach on firmer ground. It would be interesting to see how it can be related to
the work of various other groups, for example, on covariance,
\cite{DeBoer:2001uk,Cornalba:2002cu,Brecher:2004qi,Brecher:2005sj}, on higher-order terms via
stable bundles \cite{Koerber:2002zb}, and on supersymmetry
\cite{Bergshoeff:2000ik,Bergshoeff:2001dc,Refolli:2001df,Collinucci:2002ac,Drummond:2003ex}.

%%%%%%%%%%%%%%%%%%%%%%%%%%%%%%%%%%%%%%%%%%%%%%%%%%%%%%%%%%%%%%%%%%%%%%%%%%%%%%%%%%%%%%%%%%%%

\appendix

\section{Appendix}

%%%%%%%%%%%%%%%%%%%%%%%%%%%%%%%%%%%%%%%%%%%%%%%%%%%%%%%%%%%%%%%%%%%%%%%%%%%%%%%%%%%%%%%%%%%%%%%

We briefly summarise our conventions for differential forms. We use superspace conventions. A
$p$-form $\o$ on $\hM$ is written

\bea
 \o&=&\frac{1}{p!}e^{M_p}\ldots e^{M_1} \o_{M_1\ldots M_p}\nn\w2
 &:=&
 \frac{1}{p!}e^{M_p \ldots M_1}\o_{M_1\ldots M_p}
 \la{A.1}
\eea

where $e^M$ is a set of basis one-forms. The components of $\o$ are defined to be $\o_{M_1\ldots
M_p}$. The wedge product of two forms $\o,\r$, both with even overall Grassmann parity, is

\be
 (\o\wedge\r)_{M_1\ldots M_{p+q}}=\frac{(p+q)!}{p!q!}\r_{[M_1\ldots M_q} \o_{M_{q+1}\ldots
 M_{p+q}]}\ .
 \la{A.2}
\ee

The exterior derivative acts from the right by

\be
 d\o =\frac{1}{(p+1)!}e^{M_{p+1}\ldots M_1}((p+1)d_{M_1} \o_{M_2\ldots M_{p+1}} -\half
 p(p+1)f_{M_1 M_2}{}^N \o_{N M_3\ldots M_{p+1}})\ ,
 \la{A.3}
\ee

where $d_M$ denotes the vector fields dual to $e^M$ and

\be
 [d_M,d_N]=f_{MN}{}^P d_P\ .
 \la{A.4}
\ee

We have

\be
 d(\o\wedge\r)=\o\wedge d\r + (-1)^q d\o\wedge \r\ ,
 \la{A.5}
\ee

where $\r$ is a $q$-form.

A vector-valued $k$-form,

\be
 L=\frac{1}{k!} e^{M_k\ldots M_1} L_{M_1\ldots M_k}{}^N d_N\ ,
 \la{A.6}
\ee

acts as a derivation of degree $k-1$ on forms by $\o\rightarrow L\o$ where

\be
 L\o:=\frac{1}{(k+p-1)!} e^{M_{k+p-1}\ldots M_1}
 \frac{(k+p-1)!}{k!(p-1)!} L_{M_1\ldots M_k}{}^N\o_{N M_{k+1}\ldots M_{k+p-1}} \ .
 \la{A.7}
\ee

If $L$ is a vector, i.e. $k=0$, this formula reduces to the interior product of a vector with a
$p$-form.

A $(p,q)$-form with respect to some splitting of the basis set into even and odd,
$e^M=(e^m,e^{\m})$, is written

\be
 \o=\frac{1}{p!q!} e^{\m_q\ldots \m_1} e^{m_p\ldots m_1} \o_{m_1\ldots m_p \m_1\ldots \m_q}\ .
 \la{A.8}
\ee

The operation $i_N$, where $N= N^{\m\n} d_{\n}\otimes d_{\m}$, is defined by

\be
 i_N \o=\frac{1}{p!(q-2)!} e^{\m_q\ldots \m_3} e^{m_p\ldots m_1}N^{\m_2\m_1}
 \o_{m_1\ldots m_p \m_1\ldots \m_q}\ .
 \la{A.9}
\ee

If $L$ is a vector-valued one-form of the form $L=e^m L_m{}^{\m} d_{\m}$ then

\be
 L\o=\frac{1}{(p+1)!(q-1)!} e^{\m_q\ldots \m_2} e^{m_{p+1}\ldots m_1} (-1)^p(p+1)L_{m_1}{}^{\m_1}
 \o_{m_2\ldots m_{p+1} \m_1\ldots \m_q}\ .
 \la{A.10}
\ee

The sign $(-1)^p$ is explained as follows. $L$ can be viewed as a one-form which starts acting from
the right. It therefore has to be taken past $p$ $e^m s$ to act on $e^{\m_1}$. This where the sign
comes from. It also acts on the other odd basis forms giving a factor of $q$. In the text we have
used this operation with $L=\tB_1{}^1$.

\section*{Acknowledgements}

This work was supported in part by EU grant (Superstring theory) MRTN-2004-51219. UL acknowledges
partial support from VR grant 621-2003-3454.

\end{document}